\documentclass[twocolumn,prd,nofootinbib,showpacs,floatfix]{revtex4}
\usepackage{epsfig}

\newcommand{\be}{\begin{equation}}
\newcommand{\ee}{\end{equation}}
\newcommand{\bea}{\begin{eqnarray}}
\newcommand{\eea}{\end{eqnarray}}

\usepackage[dvips]{color}
\definecolor{Black}{named}{Black}
\definecolor{Red}{named}{Red}

\begin{document}

\title{Low-energy spectral features of supernova (anti)neutrinos  in inverted hierarchy}

\author{	G.~L.~Fogli$^{1,2}$, 
		E.~Lisi$^{2}$, 
		A.~Marrone$^{1,2}$,
		A.~Mirizzi$^{2,3}$ and 
		I.~Tamborra$^{1,2}$ 
\medskip\smallskip } 
\affiliation{ $^1$Dipartimento Interateneo di Fisica ``Michelangelo Merlin", Via Orabona 4, 70126   Bari, Italy  \\
		 $^2$Istituto Nazionale di Fisica Nucleare, Sezione di Bari, Via Orabona 4, 70126   Bari, Italy\\ 
		 $^3$Max-Planck-Institut f\"ur Physik (Werner-Heisenberg-Institut), F\"ohringer Ring 6, 80805 M\"unchen, Germany}


\begin{abstract}
In the dense supernova core, self-interactions may align the 
flavor polarization vectors of $\nu$ and $\overline\nu$, and induce collective 
flavor transformations. Different alignment ansatzes are known to 
describe approximately the phenomena of synchronized or bipolar oscillations, 
and the split of $\nu$ energy spectra.
We discuss another phenomenon observed in some numerical experiments in inverted
hierarchy,
showing features akin to a low-energy split of $\overline\nu$ spectra.
The phenomenon appears to be approximately described by another alignment ansatz which, 
in the considered scenario, reduces the (nonadiabatic) dynamics of all energy modes to only 
two $\nu$ plus two $\overline\nu$ modes. The associated spectral features, however, appear to be fragile when passing from single- to multi-angle
simulations. 
\end{abstract}
\pacs{ 14.60.Pq,              
       96.50.Sb,              
       95.55.Vj,
       26.65.+t \hfill Preprint MPP-2008-96}

\maketitle


\section{Introduction}

 Supernova (SN) neutrinos continue to be a subject of great interest in astroparticle
physics \cite{Raff07}. In particular, renewed attention is being paid to collective features of
flavor transformations induced by $\nu$ ($\overline\nu$) self-interactions
(see \cite{Duan08} and refs.\ therein).
The observed collective phenomena of synchronized
\cite{Past02} and bipolar \cite{Bip1,Bip2} oscillations, and
the split of $\nu$ spectra \cite{Split1,Split2} can be largely understood, in flavor space, 
in terms of various alignment (or 
antialignment) ansatzes for the Bloch polarization vectors
of $\nu$ ($\mathbf{P}$) and $\overline\nu$ ($\overline\mathbf{P}$) at different energy $E$. Conversely, lack of
alignment indicates flavor
decoherence \cite{Sigl}.

In this work we deal with a minor---yet interesting---phenomenon previously observed in the
numerical experiments of \cite{Foglilast} and then of 
\cite{Dighe,Volpe}, akin to a low-energy
``antineutrino spectral split'' in inverted hierarchy. Differently from the neutrino case, 
this feature appears to have a nonadiabatic origin. We show that the $\overline\nu$ spectral split 
observed in \cite{Foglilast} can be approximately described in terms of $\mathbf{P}$ and
$\overline\mathbf{P}$ alignment along four global modes (at low and high energy, for $\nu$ and
$\overline\nu$).
We also discuss a scenario with different input spectra, where the $\overline\nu$ split
is slightly more evident, at least in the approximation of averaged trajectories. 
The effects described herein might play a role, in principle, in future low-energy
SN neutrino observations via
$\overline\nu_e +p \to n+ e^+$ scattering.

\begin{figure}[t]
\centering
\label{fig1}
\vspace*{-7mm}
\hspace*{2mm}
\epsfig{figure=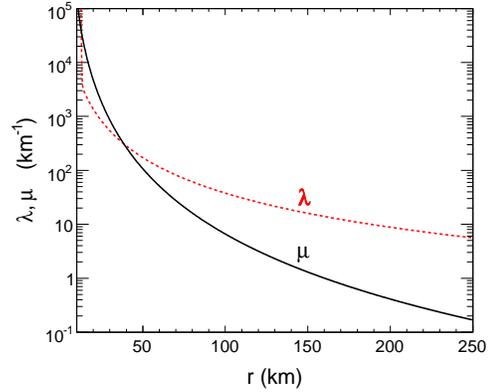,width =.87\columnwidth,angle=0}
\vspace{-4mm}
 \caption{Radial profiles adopted for the matter  ($\lambda$) and self-interaction ($\mu$) 
 potentials, in the range $r\in[10,\,250]$~km.}
\end{figure}

\vspace*{-2mm}
\section{EQUATIONS AND NUMERICAL RESULTS}
\vspace*{-1mm}

As in \cite{Foglilast}, 
we consider an effective two-flavor $(\nu_e,\nu_x)$ scenario with 
mass-mixing parameters $(\Delta m^2,\theta_{13})$. After trajectory averaging
(single-angle approximation), the $\mathbf{P}$ and $\overline\mathbf{P}$ modes obey the equations
of motion (EOM)
\begin{eqnarray}
\label{EOM1}
\dot\mathbf{P} = \mathbf{H}\times \mathbf{P} &\equiv& (+\omega \mathbf{B}+ \lambda \mathbf{z}+\mu \mathbf{D}) \times \mathbf{P}\ ,\\
\label{EOM2}
\dot{\overline\mathbf{P}} = \overline\mathbf{H}\times \overline\mathbf{P}
&\equiv & (-\omega \mathbf{B}+ \lambda \mathbf{z}+\mu \mathbf{D}) \times
\overline\mathbf{P}\ ,
\end{eqnarray}
where $\omega=\Delta m^2/2E$ is the vacuum oscillation frequency, $\lambda=\sqrt{2}G_F N_e$
is the matter potential related to the $e^-$ density ($N_e$), $\mu=\sqrt{2}G_F (N+\overline N)$
is the neutrino-neutrino potential related to the effective $\nu$ and $\overline\nu$ densities
($N$ and $\overline N$), and  $\mathbf{B} = (\sin 2 \theta_{13}, 0, \mp\cos 2 \theta_{13})$, where
the upper (lower) sign refers to normal (inverted) hierarchy. Only the latter case
will be considered hereafter.

Global polarization vectors are defined as integrals weighted by $\nu$ and $\overline\nu$
spectral densities ($n$ and $\overline n$) \cite{Foglilast}
\begin{eqnarray}
\label{J}
\mathbf{J} &=& \frac{1}{N+\overline N}\,{\int_0^\infty dE\, n\, \mathbf{P}}\ ,\\
\label{Jbar}
\overline\mathbf{J} &=& \frac{1}{N+\overline N}\,{\int_0^\infty dE\, \overline n\; \overline \mathbf{P}}\ ,
\end{eqnarray}
and their difference provides the vector $\mathbf{D}=\mathbf{J}-\overline\mathbf{J}$.

Numerically,  we take $\Delta m^2=2\times 10^{-3}$~eV$^2$, so that
\begin{equation}
\omega~[\mathrm{km}^{-1}]=\frac{5.07}{E~[\mathrm{MeV}]}\ .
\end{equation}
We also assume $\sin^2 \theta_{13}\equiv s^2_{13} = 10^{-4}$ as reference value. It is then $\mathbf B\simeq  \mathbf{z}$,
with $D_z=\mathbf{D}\cdot \mathbf{z}\simeq \mathbf{D}\cdot \mathbf{B}$ being a constant of motion \cite{Bip2}.
Concerning the
potentials $\lambda$ and $\mu$, their radial profile above the
neutrinosphere $(r>10~\mathrm{km})$ are taken as in \cite{Foglilast} and shown in Fig.~1.

\newpage

\begin{figure}[t]
\centering
\vspace*{-4mm}
\hspace*{-6mm}
\epsfig{figure=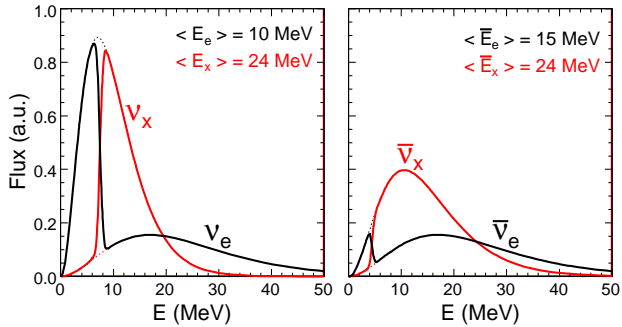,width =1.2\columnwidth}
\vspace*{-7mm}
\caption{Single-angle simulation in inverted hierarchy: 
Final fluxes (at $r=250$~km, in arbitrary units) 
for different neutrino species as a function of energy. Initial shapes ($r=10$~km) are shown
as dotted lines to guide the eye, with average energies reported on top.
\label{fig2}}
\end{figure}

Figure~2 shows typical results at the end of collective flavor transformations in inverted
hierarchy (not much happens in normal hierarchy),
as obtained by solving Eqs.~(\ref{EOM1},\,\ref{EOM2}) with the same initial conditions
as in \cite{Foglilast}. 
The most prominent feature in Fig.~2 is the $\nu$ spectral split in the left
panel at a critical energy $E_c\simeq 7$~MeV, 
but a ``minor'' $\overline\nu$ spectral split is also visible in the
right panel of Fig.~2 \cite{Note0} at $\overline{E}_c$, as observed in \cite{Foglilast}. 
A similar minor feature has been later observed as a small
low-energy ``shoulder'' \cite{Dighe} or ``bump'' 
\cite{Volpe} in the final $\overline\nu$ spectra.

\section{A DESCRIPTION WITH FOUR VECTORS}

In general, numerical solutions to Eqs.~(\ref{EOM1},\,\ref{EOM2})
can be partly understood via appropriate ``alignment'' ansatzes
for $\mathbf{P}$ and $\overline\mathbf{P}$ modes. 
In particular, 
for large $\mu$, one can posit  
that they are separately pinned to their sum
(i.e., $\mathbf{P}\parallel \mathbf{J}$ and $\overline\mathbf{P}\parallel \overline\mathbf{J}$) \cite{Past02}.
The EOM for $\mathbf{J}$ and $\overline\mathbf{J}$ can then be 
cast in a closed form, formally equivalent to the EOM of
a gyroscopic pendulum \cite{Bip2}, which elegantly explains the
so-called synchronized and bipolar oscillations observed in
numerical experiments \cite{Bip1,Bip2}.  
Alternatively, for slowly varying (adiabatic)
$\mu$, one can posit alignment with the Hamiltonians in Eqs.~(\ref{EOM1},\ref{EOM2}):
 $\mathbf{P}\parallel \mathbf{H}$ and $\overline\mathbf{P}\parallel\overline\mathbf{H}$,
explaining nicely \cite{Split1} the so-called  $\nu$ spectral split  
\cite{Split2,Smirnov,Adiab,Foglilast}.

The above two ansatzes are, in part, exclusive. The ``pendulum'' solution 
describes well the synchronized and bipolar oscillation phases,  
but not the final $\nu$ spectral split. Conversely, while the adiabatic solution describes well
overall flavor evolution features, but not the transient, nonadiabatic bipolar oscillations.
However, neither describe the low-energy $\overline\nu$ spectral split
of Fig.~2. This phenomenon must thus be related to the breaking of
global alignment and of adiabaticity. 

Adiabaticity breaking is expected in our scenario, especially
at low $E$.
According to the criterion proposed in \cite{Split1} (which is independent of 
both $\theta_{13}$ and $\lambda$), the evolution 
becomes nonadiabatic when the value of $\mu$ approaches a typical vacuum frequency $\omega$.
For $E\lesssim 40$~MeV (Fig.~2), the nonadiabaticity criterion is roughly $\mu \sim \omega \gtrsim 0.13$~km$^{-1}$, and
is satisfied at any radius $r$ in Fig.~1. Moreover, since $\mu$ is a decreasing function of
$r$, adiabaticity is violated more at high $\omega$ (low $E$). However, we have not been 
able to sharpen this qualitative expectation and to ``predict,'' for instance, the 
observed value of the $\overline\nu$ critical energy $\overline E_c$ in terms of a
specific transition to nonadiabaticity. 
For instance, we find that
$\overline{E}_c$ depends also on $\theta_{13}$ and $\lambda$; e.g., 
it decreases by $\sim\!2$~MeV by switching
off the matter term $\lambda$, but is restored to $\overline{E}_c\sim 4$~MeV for
$s^2_{13}=10^{-6}$ at $\lambda=0$. Numerical observations thus
suggest that the origin of the
$\overline{E}_c$ value is linked to 
 the complex dynamics close to the $\mathbf{z}$ axis,
which can be rather subtle.

Here we adopt a pragmatic approach, and take  $E_c$ and
$\overline{E}_c$ from the numerical experiment at face value. Our main observation is that
these energies naturally separate ``low'' ($l$) and ``high'' ($h$) energy modes, and that these
modes appear to be separately pinned to their global sum. More precisely, we define
four new global vectors by splitting the energy intervals in
Eqs.~(\ref{J},\ref{Jbar}) as $[0,E_c] \cup [E_c,\infty]$ for $\nu$
and  $[0,\overline E_c]\cup [\overline E_c,\infty]$ for $\overline\nu$,
\begin{eqnarray}
\label{Four1}
\mathbf{J}&=&\mathbf{J}_l+\mathbf{J}_h\ , \\
\label{Four2}
\overline\mathbf{J}&=&\overline\mathbf{J}_l+\overline\mathbf{J}_h\ . 
\end{eqnarray}
We find that the moduli of the above vectors ($J_l=|\mathbf{J}_l|$, etc.) are approximately conserved
during the evolution.

\begin{figure}[t]
\centering
\vspace*{-0mm}
\hspace*{-0mm}
\epsfig{figure=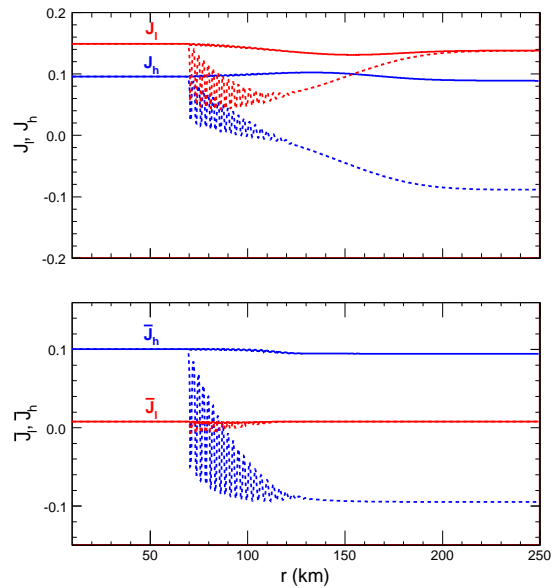,width =1.02\columnwidth}
\vspace*{-3mm}
\caption{Single-angle simulation in inverted hierarchy: Moduli and $z$-components of the
four vectors defined in Eqs.~(\ref{Four1},\ref{Four2}).
\label{fig3}}
\end{figure}

Figure~3 shows, for each of the four vectors defined above, both
the modulus and the $z$-component as a function of $r$. The initial moduli,
calculated for $E_c\simeq 7.6$~MeV and $\overline{E}_c\simeq 3.9$~MeV, are:
\begin{eqnarray}
\label{mod1}
J_l\simeq 0.149\ , && J_h \simeq 0.096\ ,\\
\label{mod2}
\overline{J}_l \simeq 0.008\ , && \overline{J}_h \simeq 0.101 \ .
\end{eqnarray}
\newpage
\noindent Such values do not change much with $r$, despite the fact
that the four vectors, initially aligned in the synchronized phase
($r\lesssim 70$~km), spread apart and nutate during
the bipolar phase ($70 \lesssim r\lesssim 120$~km) and finally split up
with low-energy (high-energy) modes aligned (antialigned) with
the $\mathbf{z}$ axis, corresponding to no (full) flavor conversion.

A posteriori, this behavior is equivalent to posit the 
(approximate) alignment 
ansatz: $\mathbf{P}\parallel\mathbf{J}_l$ for $E<E_c$, 
$\mathbf{P}\parallel\mathbf{J}_h$ for $E>E_c$, and similarly
for antineutrinos. Consistency of this ansatz with
conservation of $D_z$ from the initial (aligned) state to the final (split)
state requires
\begin{equation}
(J_l + J_h) - (\overline{J}_l+\overline{J}_h) = (J_l - J_h) - (\overline{J}_l-\overline{J}_h)\ ,
\end{equation}
namely, 
\begin{equation}
\label{equal}
J_h = \overline{J}_h \ ,
\end{equation}
which, in our case, is fulfilled within a few \%  \cite{Note1}.
The fact that the flavor evolution in Fig.~3 is reasonably described by such an ansatz  
is a clear signal of deviations from the purely adiabatic solution, which would instead
predict an alignment of the $\mathbf{P}$'s  and $\overline\mathbf{P}$'s 
to linear combinations of $\mathbf{B}$ and $\mathbf{D}$ 
\cite{Split1,Smirnov}, which we do not observe.
On the other hand, the fact that $E_c$ may change somewhat with $r$ in the
adiabatic solution \cite{Smirnov} (while we have assumed $r$-independent $E_c$ and $\overline{E}_c$ 
values a priori) suggests that our ansatz, despite its effectiveness, may 
be further refined.

Summarizing, the global flavor evolution in our scenario 
appears to be effectively described
in terms of four vectors with nearly conserved lengths, collecting
$\nu$ energy modes below and above a critical value $E_c$, and $\overline\nu$ modes
below and above a (lower) critical $\overline{E}_c$. It is then
useful to check if the behavior in Fig.~3 is also captured by
the solutions of EOM reduced to {\em exactly\/} four modes, namely,
\begin{eqnarray}
\label{first}
\dot\mathbf{J}_l &=& (+\omega_l \mathbf{B} + \lambda\mathbf{z}+ \mu \mathbf{D}) \times \mathbf{J}_l\ ,\\
\dot\mathbf{J}_h &=& (+\omega_h \mathbf{B} + \lambda\mathbf{z}+  \mu \mathbf{D}) \times \mathbf{J}_h\ ,\\
\dot{\overline{\mathbf{J}}}_l &=& (-\overline{\omega}_l \mathbf{B} +  \lambda\mathbf{z}+ \mu \mathbf{D}) \times \overline{\mathbf{J}}_l\ ,\\
\dot{\overline{\mathbf{J}}}_h &=& (-\overline{\omega}_h \mathbf{B} +  \lambda\mathbf{z}+ \mu \mathbf{D}) \times \overline{\mathbf{J}}_h\ .
\label{last}
\end{eqnarray}
where the four frequencies are defined as
averages of $\omega$ (with weights $n_e-n_x$  and $\overline{n}_e-\overline{n}_x$ for 
$\nu$ and $\overline\nu$, respectively \cite{Foglilast}) in the corresponding energy ranges ($[0,\,E_c]$ for $\omega_l$, etc.).
Numerically, in our case: $\omega_l\simeq 1.21$, $\omega_h \simeq 0.69$, 
$\overline\omega_l \simeq 2.04$, and
$\overline\omega_h \simeq 0.64$
 (all in km$^{-1}$).
The initial conditions correspond to alignment of all vectors
to $+\mathbf{z}$, with moduli given by Eqs.~(\ref{mod1},\ref{mod2}).
Such moduli are trivially conserved by construction. The $z$-components are more
interesting, and evolve in a way rather similar to those in Fig.~3  (not shown). In particular, 
since conservation of $D_z$ prevails over
exact alignment, the vectors $\mathbf{J}_l$ and $\overline{\mathbf{J}}_l$ ($\mathbf{J}_h$ and 
$\overline{\mathbf{J}}_h$) 
are almost---but not exactly---aligned (antialigned) to $+\mathbf{z}$. 
Therefore, even this simplified (4-mode) case shows
an ``antineutrino spectral split'' to a good approximation, via the opposite behavior of $\overline{\mathbf{J}}_l$ and 
$\overline{\mathbf{J}}_h$, similar to the results of the complete
(100-mode) spectral case shown in Figs.~2 and 3. 

\begin{figure}[t]
\centering
\vspace*{-2mm}
\hspace*{-0mm}
\epsfig{figure=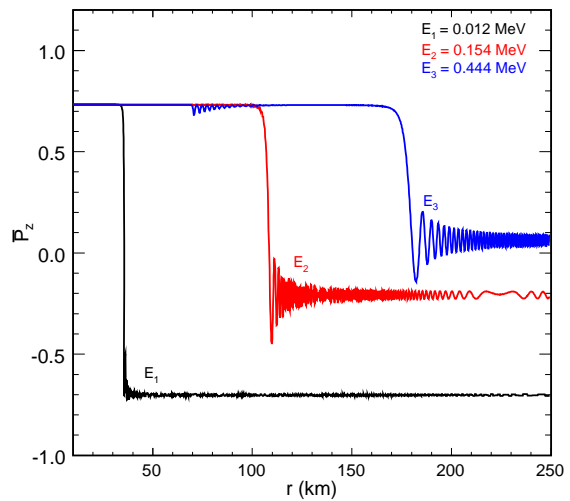,width =1.05\columnwidth}
\vspace*{-3mm}
\caption{MSW effect on the $z$-component of $\overline\nu$
modes at very low energies ($E=0.012$, 0.154, and 0.444~MeV).
\label{fig4}}
\end{figure}

We conclude by discussing the tiny effect of ordinary Mikheev-Smirnov-Wolfenstein (MSW) 
nonadiabatic resonances in matter for very low energy $\overline\nu$ modes. In our scenario
($s^2_{13}=10^{-4}$ and $\mu D_z\ll \lambda$)
the $\overline\nu$ MSW resonance
condition $\overline{\mathbf{H}}\cdot \mathbf{z}=0$ (i.e., $\omega\cos2\theta_{13}=\lambda+\mu D_z$)
approximately reads 
\begin{equation}
\label{res}
\omega\simeq \lambda\ . 
\end{equation}
This condition is met  
before the end of collective effects ($r\lesssim 250$~km) for $\omega \gtrsim 5$~km$^{-1}$ ($E\lesssim 1$~MeV),
see Fig.~1. When passing from $\lambda_i/\omega\gg 1$ (initial state $i$) to $\lambda_f/\omega\ll 1$ (final state $f$)
through the resonance, the $\overline\nu_e$ flavor
survives with probability $\overline{P}_{ee}\simeq P_c$, where
$P_c=\exp(-2\pi \omega s^2_{13}\lambda/\dot\lambda)$ is the level crossing probability \cite{KuoP}. 
The relation between $P_{ee}$ and  the $z$-components of the $\overline\nu$ polarization vectors $\overline{\mathbf{P}}$ \cite{Foglilast} 
provides the depth of the MSW resonance as
\begin{equation}
\label{depth}
\overline{P}_z^f\simeq \overline{P}_z^i(2P_c-1)\ .
\end{equation}
Figure~4 shows the radial profile of $\overline{P}_z$, as obtained
in our numerical experiment for three representative (very low) energies. 
The radius and depth of the MSW resonance agree with Eqs.~(\ref{res})
and (\ref{depth}), respectively. The lower the energy, the deeper the resonance,
the more inverted is the final polarization vector $\overline P$. However, the non-collective
($\omega>\mu$) MSW effect is limited to such low energies to be practically
unobservable in Fig.~2, and is numerically irrelevant for the 4-mode approximation described above. 
Finally, we mention that, in {\em normal\/} hierarchy, we find that the spectra of $\nu$ and $\overline\nu$
are basically unaffected by self-interaction effects, the only tiny change being a spectral swap 
of {\em neutrinos\/} for $E \lesssim 1$~MeV, induced by MSW  effects (not shown).

\newpage
\begin{figure}[t]
\centering
\vspace*{-4mm}
\hspace*{-6mm}
\epsfig{figure=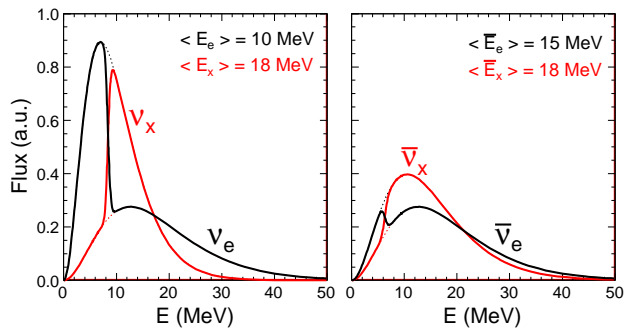,width =1.2\columnwidth}
\vspace*{-8mm}
\caption{As in Fig.~2 (single angle), but for $\langle E_x\rangle=18$~MeV.
\label{fig5}}
\end{figure}
\vspace*{-2mm}

\vspace*{-10mm}
\section{SINGLE- VERSUS MULTI-ANGLE}
\vspace*{-1mm}

So far we have examined in detail the low $E$ features of the same
scenario of \cite{Foglilast} in single-angle approximation. It was shown in
\cite{Foglilast}, however, that
the features in Fig.~2 (most notably the $\overline\nu $ spectral split) may be smeared out in more realistic (``multi-angle'' \cite{Duan08}) numerical experiments accounting for different (unaveraged) trajectory angles. 

Here we study the transition from single- to multi-angle  
simulations in a (hypothetically) more favorable
scenario, where $\langle E_x\rangle$ is decreased from 
24 MeV to 18 MeV---everything else being unchanged. This choice,
due to a ``fatter'' low-energy tail, leads to a slightly more prominent $\overline\nu$ split
at higher-$\overline{E}_c$, as evident in the single-angle results
of Fig.~5 (to be compared with Fig.~2).

Concerning the reduction to four modes, we find that
the case in Fig.~5 is well described 
by $E_c\simeq 8.2$~MeV,  
$\overline E_c\simeq 6.1$~MeV, and:
$J_l\simeq 0.131$, $J_h \simeq 0.031$,
$\overline{J}_l \simeq 0.012$,  $\overline{J}_h \simeq 0.029$.
The corresponding frequencies  
are: $\omega_l\simeq 1.16$, $\omega_h \simeq 0.97$, 
$\overline\omega_l \simeq 1.36$, and
$\overline\omega_h \simeq 0.66$ (in km$^{-1}$).

Concerning the transition from single- to multi-angle calculation,
however, we find again that the latter tend to smear out the low-energy
spectral features. Figure~6 (multi-angle) shows, in comparison with Fig.~2 (single-angle) that: (1) the $\nu$ split is broadened, and (2) the $\overline \nu$ one
is largely suppressed, and survives as a slight low-energy ``shoulder.'' 
At present, this (unfavorable) numerical 
observation \cite{Note2} remains analytically unexplained. One cannot exclude, however,
that the $\overline\nu$ spectral split feature may be less suppressed in other
 multi-angle cases, whose investigation is left to future work.

\vspace*{-4mm}
\section{Summary}
\vspace*{-2mm}

We have studied some low-energy features of the
$\overline\nu$ energy spectrum, focusing on the spectral split
phenomenon emerging (in inverted hierarchy)
in single-angle simulations. We have related its origin to nonadiabatic
aspects of the $\nu$ and $\overline\nu$ evolution, which can be simply 
modeled through an alignment ansatz with four energy modes. However,
the $\overline\nu$ split feature appears to be fragile when passing to
multi-angle simulations. Further studies are required to deepen its
analytical understanding, as well as the conditions for its 
observability.

\begin{figure}[b]
\centering
\vspace*{-7mm}
\hspace*{-6mm}
\epsfig{figure=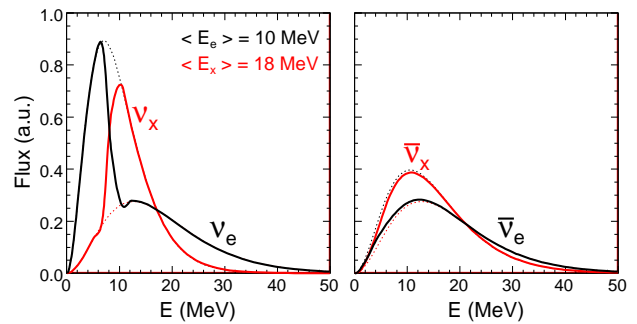,width =1.2\columnwidth}
\vspace*{-8mm}
\caption{As in Fig.~5, but in multi-angle simulation.
\label{fig6}}
\end{figure}

\vspace*{1mm}
{\bf Acknowledgments.}
This work is supported in part by
the Italian INFN and MIUR (Astroparticle Physics project),
and by the E.U.\ (ENTApP network). 
In Munich, A.M.\ is supported by an INFN postdoc fellowship.

\vspace*{-2mm}

\end{document}